\documentclass[preprint]{elsarticle}

\usepackage{subfigure}
\usepackage{amssymb}
\usepackage{amsmath}

\newcounter{bla}

\journal{Computer Physics Communication}

\begin{document}

\begin{frontmatter}
\title{A program for the Bayesian Neural Network in the ROOT framework}

\author[nju,ipas]{Jiahang Zhong}
\author[nju]{Run-Sheng Huang}
\author[ipas]{Shih-Chang Lee\corref{cor1}}
\cortext[cor1]{Corresponding author. \\\textit{E-mail address:} phsclee@phys.sinica.edu.tw}
\address[nju]{School of Physics, Nanjing University, CN - Nanjing 210093, China}
\address[ipas]{Institute of Physics, Academia Sinica, TW - Taipei 11529, Taiwan}

\begin{abstract}
We present a Bayesian Neural Network algorithm implemented in the TMVA package~\cite{TMVA}, within the ROOT framework~\cite{ROOT}. Comparing to the conventional utilization of Neural Network as discriminator, this new implementation has more advantages as a non-parametric regression tool, particularly for fitting probabilities. It provides functionalities including cost function selection, complexity control and uncertainty estimation. An example of such application in High Energy Physics is shown. The algorithm is available with ROOT release later than 5.29.
\end{abstract}


\end{frontmatter}

{\bf PROGRAM SUMMARY}\\
\begin{small}
\noindent
{\em Manuscript Title:} A program for the Bayesian Neural Network in the ROOT framework                                       \\
{\em Authors:} Jiahang Zhong, Run-Sheng Huang, Shih-Chang Lee                                               \\
{\em Program Title:}TMVA-BNN                                  \\
{\em Journal Reference:}                                      \\
{\em Catalogue identifier:}                                   \\
{\em Licensing provisions:} BSD                                  \\
{\em Programming language:} C++                                   \\
{\em Computer:} Any computer system or cluster with C++ compiler and UNIX-like operating system.                                               \\
{\em Operating system: }Most UNIX/Linux systems. The application programs were thoroughly tested under Fedora and Scientific Linux CERN. 
{\em Classification:} 11.9                                        \\
{\em External routines/libraries:} ROOT package version 5.29 or higher (http://root.cern.ch)                \\
{\em Nature of problem:} Non-parametric fitting of multivariate distributions.\\
   \\
{\em Solution method:} An implementation of Neural Network following the Bayesian statistical interpretation. Uses Laplace approximation for the Bayesian marginalizations. Provides the functionalities of automatic complexity control and uncertainty estimation.\\
   \\
{\em Running time:} Time consumption for the training depends substantially on the size of input sample, the NN topology, the number of training iterations, etc. For the example in this manuscript, about 7 minutes was used on a PC/Linux with 2.0GHz processors. \\
   \\
\end{small}

\section{Introduction}
\label{sec:Intro}

\begin{figure}[htbp]
\begin{center}
\includegraphics[width=0.6\textwidth]{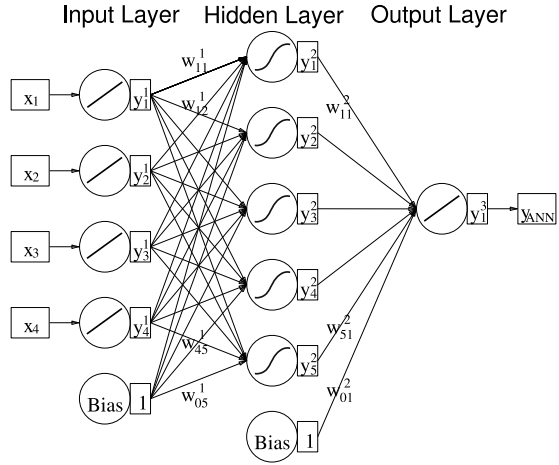}
\caption{The structure of an NN with one hidden layer~\cite{TMVA}.}
\label{fig:MLPstructure}
\end{center}
\end{figure}

Neural Network (NN) has been considerably utilized in High Energy Physics in the past decade. In most applications, The NN was used as a discriminator to separate signal from backgrounds. It is a powerful tool to extract the features of target categories in the multivariate phase space, and project them into a scalar discriminator. However, such applications are often criticized for the reliance on the simulation or test-beam data as training samples, which may have distinct features comparing to the real data. On the other hand, the usage of NN as a non-parametric regression tool is much less exploited, and usually does not suffer from such concerns. Given sufficient complexity, even a single-hidden-layer NN can be seen as a universal approximator of any nonlinear multivariate function~\cite{Cybenko:1989,Hornik1989359}. Composed of simple nonlinear functions called ``neurons", such as hyperbolic tangent functions, an NN can achieve great complexity by connecting many neurons with variable weights $\mathbf{w}$, which serves as the free parameters of the model. Equation~(\ref{eqn:NNfunction}) shows the analytical form of a typical NN, with its structure shown in figure~\ref{fig:MLPstructure}.

\begin{equation} 
\label{eqn:NNfunction}
\begin{split}
&y_i^1=\vphantom{\sum_i} f^{(1)}(x_i)  \qquad\qquad\qquad\quad f^{(1)}(h)=ah+b  \\
&y_j^2=f^{(2)}(w_{0j}^1+\sum_i w_{ij}^1 y_i^1)  \quad f^{(2)}(h) = \tanh(h)  \\
&y_1^3=f^{(3)} (w_{01}^2+\sum_i w_{j1}^2 y_j^2)  \quad f^{(3)}(h) = h 
\end{split}
\end{equation}

NN can approximate not only functions whose output values span real number space, but also those with confined output values. One such category particularly interesting is a probability. The output value can be confined within 0 and 1 by applying a sigmoid transformation to the output neuron, i.e. replacing $f^{(3)}$ in equation~(\ref{eqn:NNfunction}) by
\begin{equation}
\label{eqn:sigmoid}
\begin{split}
&f^{(3)}(h) = \frac{1}{1+e^{-h}}
\end{split}
\end{equation}

There are several advantages of NN comparing to the conventional representation of probabilities by histograms. First, NN can approximate the distributions in an unbinned manner, without the arbitrariness in the choice of binning and the subsequent loss of information. Second, NN is more practical for multivariate approximation without suffering the curse of dimensionality. Another great advantage of NN as a regression tool is that the correlations between the input variables can be well approximated.

Compared to the application as a ``black-box" discriminator, NN application as a regression tool must follow a more explicit statistical interpretation. This is particularly important if the subsequent application requires detailed statistical information, such as limit setting for new physics searches. Here we present an NN algorithm following the Bayesian inference theory, implemented in the TMVA package~\cite{TMVA}. This manuscript is organized as follows: section~\ref{sec:Bayesian} is a brief review of the statistical interpretation for the training (fitting) and prediction procedure of the NN. Then the implementation of our Bayesian NN (BNN) algorithm is described in detail in section~\ref{sec:Implementations}. Finally in section~\ref{sec:Application}, an application of this BNN in High Energy Physics is demonstrated, where it is used to approximate the false identification rate of isolated muons.

\section{Statistical Interpretation of Training and Prediction}
\label{sec:Bayesian}

As a generic model, NN normally has a huge number of degrees of freedom and incomprehensible complexity. On the other hand, the procedure to train an NN and makes predictions with it can be clearly interpreted by probability theory as Bayesian inference. 

Given an observed sample $\mathbf{D}=\{\mathbf{x_i}\,, t_i \}$, with $\mathbf{x_i}$ as the multiple input variables of entry $i$, and $t_i$ as its observed output value, the training (fitting) of NN can be viewed as a process to determine the probability of free parameters $\mathbf{w}$, based on $\mathbf{D}$. According to the Bayes theorem, this ``posterior" probability comes from the combination of previous knowledge of $\mathbf{w}$ (``prior") and the compatibility of sample $\mathbf{D}$ with the NN (``likelihood")
\begin{equation} 
\label{eqn:Training}
\begin{split}
P(\mathbf{w}|\mathbf{D}) \propto P(\mathbf{D}|\mathbf{w}) P(\mathbf{w})
\end{split}
\end{equation}

The likelihood $P(\mathbf{D}|\mathbf{w})$ in Bayesian language is closely related the ``cost function" in machine learning language. The optimization of NN by minimizing the cost function is mostly equivalent to maximizing the likelihood. For example, the most commonly used sum of the square error (SSE) function can actually be translated as the negative logarithm of the Gaussian likelihood, as shown by equation~(\ref{eqn:SSE}). 
\begin{equation} 
\label{eqn:SSE}
\begin{split}
\text{SSE}&=\sum_i ( y(\mathbf{x_i;\mathbf{w}})-t_i )^2 \\
&= -\log(\prod_i (\exp(-(y(\mathbf{x_i};\mathbf{w})-t_i)^2)\\
&\propto -\log(P(\mathbf{D}|\mathbf{w}))
\end{split}
\end{equation}

The prior probability $P(\mathbf{w})$ is much less emphasized in classic usage of NN, which in fact often assumes a flat distribution. We will see later that a Bayesian regulator term can be added to the cost function, based on a simple prior knowledge. It is also worth mentioning that, although the probability distribution of $\mathbf{w}$ can be obtained, only the most probable value is kept in classic usage.

Given a new input vector $\mathbf{x'}$, the prediction with NN can also be performed as a Bayesian inference. Although classic usage of NN only gives one single value of $y'$ with the most probable $\mathbf{w}$, we should be able to predict a probability distribution of the output if we marginalize over the distribution of NN parameters $P(\mathbf{w|D})$ by equation~(\ref{eqn:Prediction}).
\begin{equation} 
\label{eqn:Prediction}
\begin{split}
& P(y'| D, \mathbf{x'} ) = \int P(y'|\mathbf{x'}\,,\mathbf{w}) P(\mathbf{w}| D) \, d\mathbf{w}
\end{split}
\end{equation}

\section{Bayesian Implementations}
\label{sec:Implementations}

\subsection{Cost Function}
\label{sec:CostFunction}

As shown in equation~(\ref{eqn:SSE}), the commonly used cost function, sum of the square error, can be interpreted as the negative logarithm of Gaussian likelihood function. This cost function is applicable for most regression applications, where a Gaussian distribution can be assumed for the observed values around their true values. 

However, when the NN is used to approximate a probabilistic distribution, such an assumption of Gaussian likelihood may not be appropriate. As a probabilistic classifier, the NN's output $y$ is expected to approximate the probability of membership to one category, constrained between 0 and 1 by the sigmoid function (equation~(\ref{eqn:sigmoid})). And the observed value $t$ for each entry in the input sample $D$ is usually either 0 or 1, representing the fact whether the entry meets the condition, or belongs to the desired category. The distribution of observation $t$ around probability $y$ should then follow the Bernoulli distribution 
\begin{equation} 
\label{eqn:BernL}
\begin{split}
&P(\mathbf{D}|\mathbf{w})=y^t(1-y)^{1-t}
\end{split}
\end{equation}
Correspondingly, the cost function for classification should take the form of the so-called ``cross-entropy" function (CE), the sum of negative logarithm of Bernoulli distribution, 
\begin{equation} 
\label{eqn:CE}
\begin{split}
\text{CE}&=\sum_i (-t_i\log y(\mathbf{x_i};\mathbf{w})-(1-t_i)\log(1-y(\mathbf{x_i};\mathbf{w})))
\end{split}
\end{equation}
With the sigmoid transformation and CE cost function, the NN can approximate the probabilities with rigorous statistics interpretation. This can be chosen by the option of the MLP method~\cite{TMVA} {\tt EstimatorType=CE}. 

\subsection{Complexity Control}
\label{ComplexityControl}

NN gains the capability of universal approximation by a huge number of degrees of freedom. A typical single-hidden-layer NN, even only for a few input variables, could have $O(10^2)$ free parameters $\mathbf{w}$. A model with such great complexity may suffer from over-fitting. That is to say, it will approximate not only the desired connection between the input variables and the output value, but also the undesired fluctuations of the input sample. This is particularly an issue if the input sample has limited statistics. The predictivity of the model will be greatly deteriorated by exaggerated fluctuations. 

The commonly employed solution against over-fitting is the so-called ``cross validation" technique. A fraction of the input sample, normally half of the statistics, is taken away from the training process and used as a test set. Over-fitting is identified during the process of training, if the cost function of the test set starts to increase. One problem with this technique is the possibility that the cost function of test set may have merely hit a local minimum. Another problem is the reduction of training sample size. This is a non-trivial drawback for cases in which over-fitting may occur, which normally have trouble with statistics already.

In our implementation, another solution with regulators is adopted to avoid over-fitting. Although it is necessary to keep a large number of free parameters in order to make the model generic, the value of the parameters can be constrained to reduce unnecessary complexity. This can be expressed as a Bayesian prior knowledge about the model, assuming the value of the free parameters $\mathbf{w}$ should be limited to the vicinity of zero. A Gaussian distribution centered at zero is used to represent such prior. Correspondingly, a ``regulator" function can be obtained as the negative logarithm of this Gaussian prior for all NN parameters $w_i$.
\begin{equation} 
\label{eqn:regulator}
\begin{split}
&\text{Reg}=-\log(P(\mathbf{w}))=\sum_i  (\alpha_i w_i^2) 
\end{split}
\end{equation}
Adding this regulator term into the cost function actually gives the negative logarithm of posterior probability $P(\mathbf{w}|D)$. In the BNN, this summed value is minimized instead to obtain the optimal $\mathbf{w}$.

In equation~(\ref{eqn:regulator}), the hyper-parameters $\alpha_i$ determine the range of $\mathbf{w}$, consequently reflecting the knowledge of required complexity of the model. The values of $\alpha_i$ are not necessarily the same for each $w_i$. From a topological point of view, all the neurons within one hidden layer are computationally exchangeable. So their outgoing weights share the same hyper-parameter. This is not applicable to the input variables because they normally have different importance. Therefore, each group of weights originated from the same input neuron has its own $\alpha_i$. For the same reason, the bias neurons in each layer have independent hyper-parameters.

Although we can categorize the weights and associate them to different hyper-parameters, in most cases we do not actually possess \textit{a priori} knowledge about the complexity needed, namely the values of $\alpha_i$. We implemented an iterative approach proposed by MacKay in 1992~\cite{MacKay:1992}, in which these hyper-parameters are estimated during the training of NN, by optimizing the ``evidence" of the models:
\begin{equation} 
\label{eqn:evidence}
\begin{split}
P( D | \pmb{\alpha} ) & = \int P( D | \mathbf{w},\pmb{\alpha}) P(\mathbf{w} | \pmb{\alpha}) \, d\mathbf{w}
\end{split}
\end{equation}
where the optimal $\mathbf{w}$ is determined by minimizing the cost function and the integral is evaluated approximately for a given $\pmb{\alpha}$. After estimating the optimal $\pmb{\alpha}$, a new optimal $\mathbf{w}$ is recalculated and the process is iterated, until desired convergence is reached.

This functionality can be activated by the MLP option {\tt UseRegulator=true}, together with the BFGS training method.

\subsection{Bayesian Prediction}
\label{sec:BayesianPrediction}

In the classic usage of NN, the prediction upon new input $\mathbf{x'}$ is the most probable value obtained with the most probable $\mathbf{w}$. Instead, in Bayesian data analysis, it is more essential to marginalize over all possible values of $\mathbf{w}$, and obtain the prediction as a probability distribution, as shown by equation~(\ref{eqn:Prediction}). Such a prediction contains not only the most probable value, but also the uncertainty of the inference. The estimation of uncertainty is crucial, especially for extrapolated predictions in multivariate phase space.

Unfortunately, as a common difficulty for most Bayesian applications, the integration of equation~(\ref{eqn:Prediction}) is generally non-trivial. Although the posterior of NN parameters $P(\mathbf{w}|D)$ is calculable for any $\mathbf{w}$, the distribution for such ``nuisance parameters" does not have a closed-form expression. In our implementation, an analytical distribution is used for the integration, which is the Laplace approximation of the posterior around the optimal value $\mathbf{w^\text{MP}}$~\cite{Mackay:1995}, as shown in equation~(\ref{eqn:GausApp}).
\begin{equation} 
\label{eqn:GausApp}
\begin{split}
& P(\mathbf{w}| D) \simeq P(\mathbf{w^{\text{MP}}}|D) \exp\left(-\frac{1}{2}\Delta \mathbf{w^{\text{T}}}  \mathbf{A}  \Delta \mathbf{w}\right)  \\
&\Delta \mathbf{w=w-w^{\text{MP}}}
\end{split}
\end{equation}
$\mathbf{A}$ represents the Hessian matrix of the cost function, namely the negative logarithm of the posterior
\begin{equation} 
\label{eqn:Hessian}
\begin{split}
\mathbf{A}&=-\nabla\nabla\log P(\mathbf{w}|D) | _{\mathbf{w}^{\text{MP}}}.  \\
\end{split}
\end{equation}

For feed-forward NN either with a linear output neuron and SSE cost function, or with a sigmoid output neuron and CE cost function, its Hessian matrix $\mathbf{A}$ can be approximated and written consistently as  
\begin{equation} 
\label{eqn:ApproximateHessian}
\begin{split}
\mathbf{A}&\simeq\sum_{\mathbf{x_i}}f'\mathbf{g^Tg}
\end{split}
\end{equation}
$f'=dy/dh$ and $\mathbf{g}=\nabla h$. $h$ and $y$ are the values before and after output neuron transformation. 

Furthermore, a linear dependence of $h$ over weights $\mathbf{w}$ can be approximated as well. 
\begin{equation} 
\label{eqn:LinApp}
\begin{split}
& h( \mathbf{x'} ; \mathbf{w} )  \simeq h(  \mathbf{x'} ; \mathbf{w^{\text{MP}}} ) + \mathbf{g} \cdot\Delta \mathbf{w}
\end{split}
\end{equation}
Therefore, the distribution of $h$ can be calculated analytically as 
\begin{equation} 
\label{eqn:ApproximatePrediction}
\begin{split}
P(h| D, \mathbf{x'} ) &= \int h(\mathbf{x'};\mathbf{w}) P(\mathbf{w}| D) \, d\mathbf{w}\\
&\simeq \mathcal{N}(\, h(\mathbf{x'};\mathbf{w^{\text{MP}}})\,,\mathbf{g^TA^{-1}g})
\end{split}
\end{equation}
It is a Gaussian distribution with the mean value of the classic NN prediction. In addition, each prediction can give an associated uncertainty, which originates from the uncertainty of the NN parameters determination. For probability fitting, the probability density function (p.d.f.) of the output $y$ is a sigmoid-transformed Gaussian distribution of $h$ (equation~\ref{eqn:sigmoid}). As a non-linear transformation, the sigmoid function will convert the symmetric error bar of h into an asymmetric one for y.

Besides the Laplace approximation used in our implementation, the integration can also be solved numerically by Markov Chain Monte Carlo (MCMC)~\cite{Neal:1993}. Compared to MCMC approach, the analytical approximation is easier for both training and prediction. The estimation of the Hessian matrix during the training process can be activated by MLP option {\tt CalculateErrors=true}. By configuring the Reader option {\tt Error=true}, the Hessian matrix will be loaded for prediction. And the asymmetric uncertainties can be evaluated by Reader functions {\tt GetMVAErrorUpper()} and {\tt GetMVAErrorLower()}.

\section{Application in High Energy Physics}
\label{sec:Application}

To demonstrate the practical usage of the BNN, we will show an example in High Energy Physics: measurement of the false identification rate of isolated muons. The test data, job control scripts and instructions can be obtained from the CPC library.

Muons are an important electroweak signature in collider physics. According to their sources, they can be categorized by the so-called isolation condition, i.e. adjacent particle flow. Those ``non-isolated" muons, which have accompanying particles flying in a similar direction, mostly come from semi-leptonic decay of heavy flavor quarks ($b,c$). The other ``isolated" muons are more likely from electro-weak processes such as $W$ boson production, and therefore taken as signals in many physics topics. To get rid of the former category from a mixed sample, an isolation cut is often imposed on the sum of transverse momenta, for all visible particles inside a cone around the muon. Here we choose a typical configuration, limiting transverse particle flow within $\Delta R=\sqrt{(\Delta\eta)^2+(\Delta\phi)^2}\leq0.2$ to be less than 15\% of the muon's transverse momentum.

Due to event-specific kinematics and detector response, there is always a considerable fraction of muons from ``non-isolated sources" being falsely identified as isolated muons. Such muons will contaminate the signal sample when a final state with isolated muons is expected. It is non-trivial to estimate this background accurately, for both new physics searches and precise measurements. Rather than relying on simulation, it is highly desirable to measure the false identification rate $f=P(\text{isol})$ directly from collision data. 

In the following, we demonstrate how BNN can be used for such false rate measurement, using Monte-Carlo simulation of heavy-flavor quark production ($b\bar{b},c\bar{c}$) in the proton-proton collisions. The samples are produced by the Pythia generator~\cite{Pythia}, with 7 TeV center-of-mass energy as at LHC~\cite{LHC}. Muon pseudo-rapidity acceptance is assumed to be within 2.5, with the threshold of transverse momentum as above 10 GeV. The fiducial efficiency of the detector is assumed to be 100\% for simplicity. 

Two samples are generated for comparison. The first sample requires single muon within the detector acceptance, while the second sample requires two such muons with the same charge. The former final state is dominated by the non-isolated sources, and therefore ideal for the measurement of fake rate. The latter final state instead is a typical channel in which new physics may manifest, and requires accurate estimation of the background contribution. With collision data, we need to measure the fake rate from the single muon events, then apply it to those same-charge di-muon events to estimate the background contribution~\cite{SSpaper}. With the MC samples mentioned above, we can study how to make the fake rate measurement compatible between these two channels. 

\begin{figure}[htbp]
\begin{center}
  \includegraphics[width=0.4\textwidth]{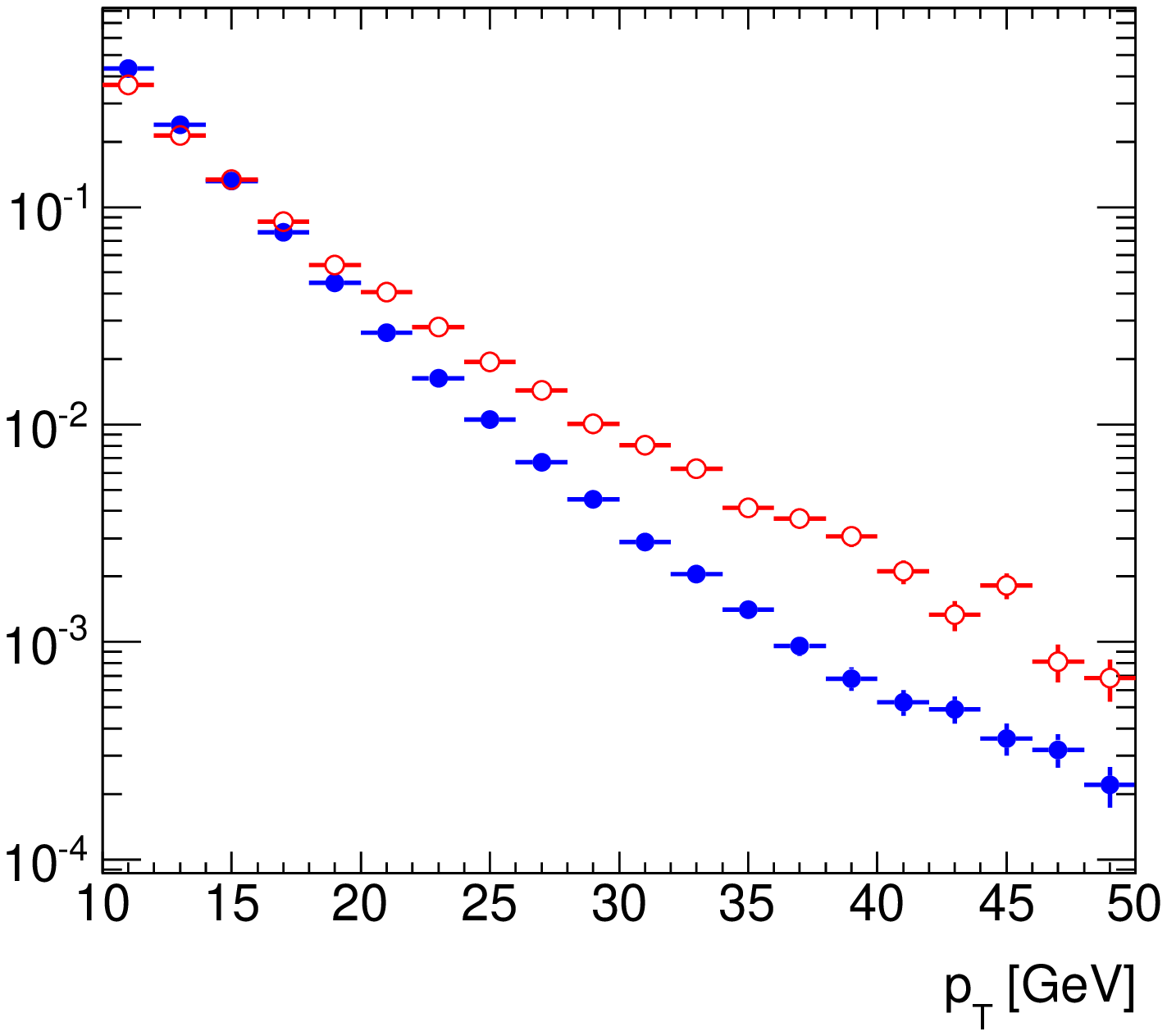}
  \includegraphics[width=0.4\textwidth]{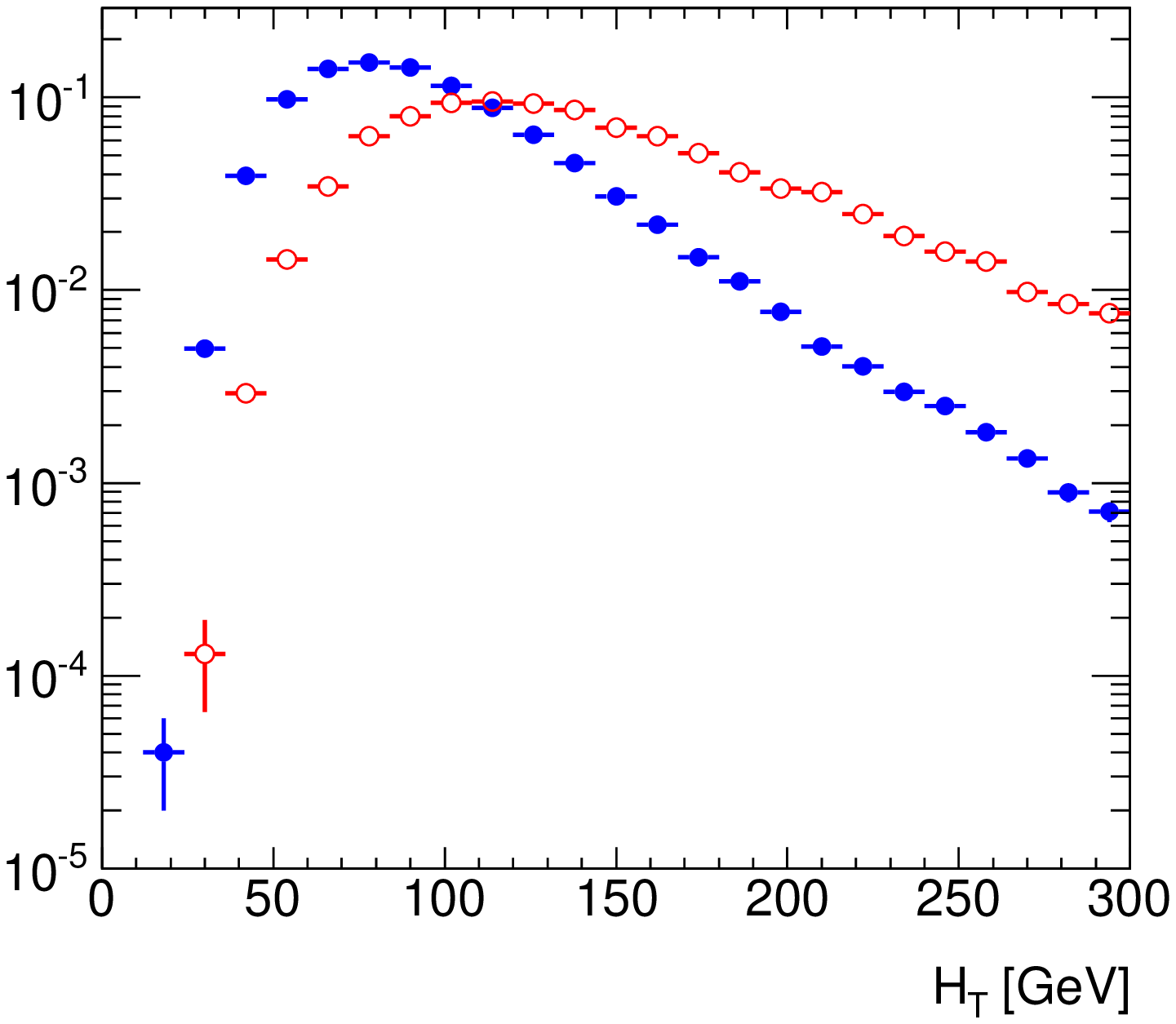}
\caption{Probability density distributions over $p_{\text{T}}$ (left) and $H_{\text{T}}$ (right), for muons in the single muon control sample (solid) and di-muon signal sample (circle).}
\label{fig:Distdependence}
\end{center}
\end{figure}

In a first attempt to measure the average false identification rates, we obtain a quite different values, 33.3(0.1)$\pm$\% from the single muon sample and 19.8(0.2)\% from the same-charge di-muon sample. This incompatibility is due to the fact that the probability of false identification, $P(\text{isol}|\mathbf{x})$, is not constant over kinematic variables $\mathbf{x}$. Two examples of such kinematic variables are
\begin{itemize}
\item $p_{\text{T}}$: the transverse momentum of the muon itself.
\item $H_{\text{T}}$: the scalar sum of transverse momenta for all visible particles in the event, and the measurable energy imbalance.
\end{itemize}
The kinematic distributions in these two samples are quite distinct, as can be seen in figure~\ref{fig:Distdependence}. As a result, the observed average rates, marginalized by equation~(\ref{eqn:AvgK}), turned out to be incompatible. In order to make a correct prediction in the signal region, it is important to measure the probabilities $P(\text{isol}\,|\,\mathbf{x})$ rather than the marginalized rate. 
\begin{equation}
\label{eqn:AvgK}
\begin{split}
&\langle f\rangle=\int P(\text{isol}\,|\,\mathbf{x})P(\mathbf{x})\,d\mathbf{x}
\end{split}
\end{equation}

Our BNN implementation is used to perform this measurement of $P(\text{isol}\,|\,\mathbf{x})$. About 50,000 muons in the single muon QCD events is used as input sample $\mathbf{D}$. Their corresponding kinematic variables $p_{\text{T}}$ and $H_{\text{T}}$ are declared as the input $\mathbf{x_i}$ of the NN. And a target value $t_i$ as 1 or 0 is assigned, depending on whether the muon passed the isolation criteria. The NN is constructed with one hidden layer of ten neurons, and a sigmoid-transformed output neuron. As a probability fit, the training is configured to use the cross-entropy cost function of equation~(\ref{eqn:CE}). In addition, the regulator mechanism is activated to prevent over-fitting.

\begin{figure}[htbp]
\begin{center}
  \includegraphics[width=0.65\textwidth]{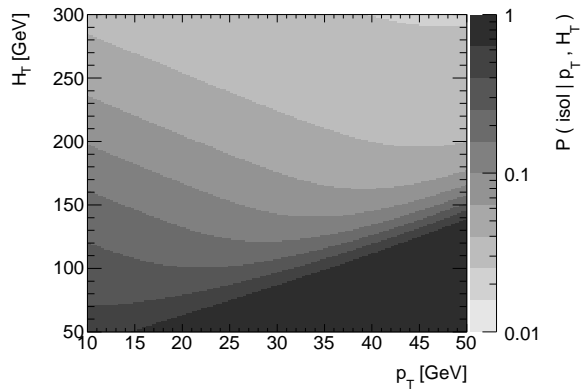}
\caption{2-dimensional false isolation probabilities, fitted by the BNN.}
\label{fig:BNNmap}
\end{center}
\end{figure}

The fitted distribution $P(\text{isol}\,|\,p_{\text{T}},H_{\text{T}})$ can be visualized in figure~\ref{fig:BNNmap}. The correlation between the two variables is clearly fitted, with a smooth extrapolation into the peripheral phase space region. 

We then test this 2D false rate function with muons in the same-charge di-muon sample. Using the trained BNN, we can predict the probability of passing isolation criteria $P(\text{isol}\,|\,p_{\text{T}},H_{\text{T}})$ for each muon. Marginalizing the predicted probabilities over all the muons in this sample (30764 muons), we can predict that 6077 of them will pass the isolation cut, corresponding to an average false rate of 19.8\%. This is very close to the actual number of passed muon, 6093, equivalent to an average false rate of 19.8(0.2)\%.

\begin{figure}[htbp]
\begin{center}
   \subfigure[]{
     \label{fig:SigmaRatio}
     \includegraphics[width=0.45\textwidth]{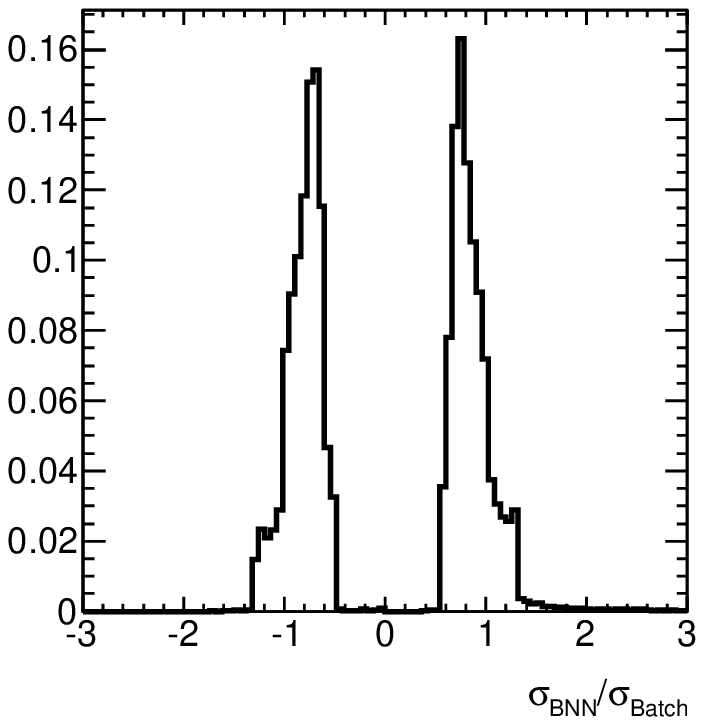}
  }
  \subfigure[]{
     \label{fig:BNNvsBatch}
     \includegraphics[width=0.45\textwidth]{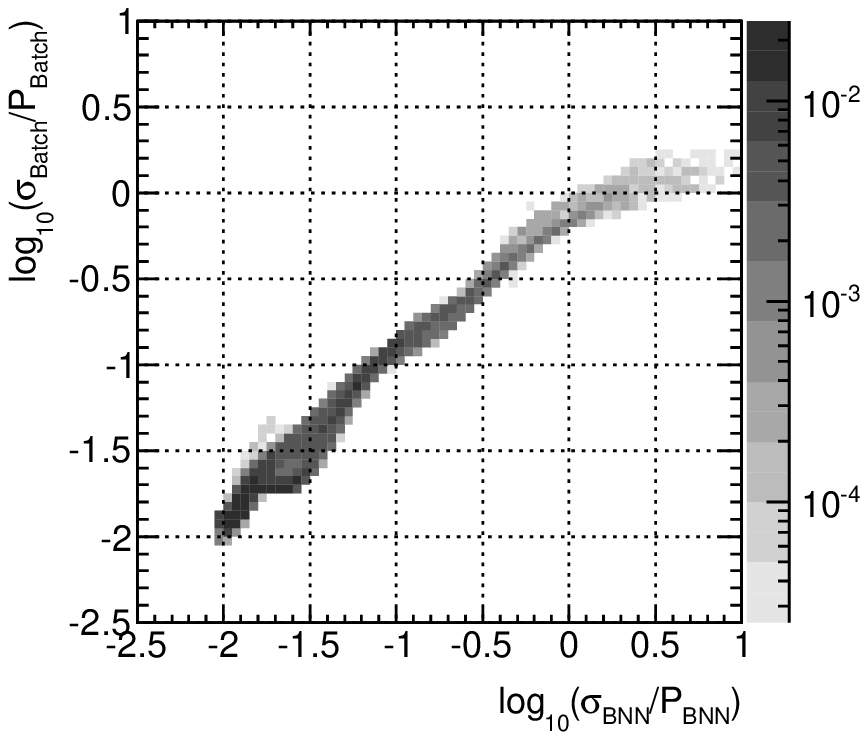}
  }
\caption{(a) The distribution of the ratios between $\sigma_\text{BNN}$ and $\sigma_\text{Batch}$, for all muons in the same-charge di-muon events. Positive value stands for $\sigma_\text{BNN}^+/\sigma_\text{Batch}$ and negative value stands for $\sigma_\text{BNN}^-/\sigma_\text{Batch}$. (b) The correlation between $\log_{10}(\sigma_\text{BNN}/P_\text{BNN})$ and $\log_{10}(\sigma_\text{Batch}/P_\text{Batch})$. $\sigma_\text{BNN}$ is the average of $\sigma_\text{BNN}^+$ and $\sigma_\text{BNN}^-$. }
\label{fig:Uncertainties}
\end{center}
\end{figure}

As described is section~\ref{sec:BayesianPrediction}, the BNN can also calculate the uncertainty associated to each prediction, based on the uncertainty on the determination of the free parameters $\mathbf{w}$. It reflects the statistical property of the training sample. To test this estimation, another 100 NNs are trained with the same network topology. But the input samples, as well as the random seeds for initialization, are different in each training. For every muon in the same-charge di-muon sample, we use the BNN to predict its probability of pass $P_\text{BNN}$, as well as the associated asymmetric error bars, denoted as $\sigma_\text{BNN}^+$ and $\sigma_\text{BNN}^-$. As a comparison, we also use the ``batch'' NNs to make 100 predictions, and calculate their mean value $P_\text{Batch}$ and standard deviation $\sigma_\text{Batch}$. Comparing the ratio between $\sigma_\text{BNN}$ and $\sigma_\text{Batch}$ (figure~\ref{fig:SigmaRatio}) for all the muons, we can see that the BNN uncertainty is generally consistent with the standard deviation of the batch predictions. 

To further understand the precision of the BNN uncertainty estimation, the correlation between $\log_{10}(\sigma_\text{BNN}/P_\text{BNN})$ and $\log_{10}(\sigma_\text{Batch}/P_\text{Batch})$ for all the muons are plotted in figure~\ref{fig:BNNvsBatch}. Good consistency can be observed when the relative uncertainties are less than $\sim$30\%, which is the case for a large fraction of the entries. For predictions with large relative uncertainty, BNN tends to over-estimate the uncertainty, as the approximation applied in the Bayesian marginalization becomes less accurate.
  
It is worthwhile to notice that the uncertainty estimated by BNN only accounts for the confidence in the determination of BNN parameters, evaluated based on the training sample used. The difference between the prediction and observation also involves the statistical fluctuations of the prediction sample, as well as the systematic uncertainties in the application, such as sample selections, the choice of parameterized variables and their measurement uncertainties. 

Furthermore, only two kinematic variables were considered in this truth-level study, and consistent false rate estimation has already been observed. In reality, there could be additional factors which considerably affect the false rate, due to detector effects and collision configuration. Fortunately, the measurement can be easily extended to a higher dimensional phase space with BNN as the fitting tool.
	
\section{Conclusion}
\label{sec:Conclusion}

In this manuscript we presented a BNN algorithm which can be used as an unbinned fitting tool, which is particularly interesting for fitting probabilities. The Bayesian implementation also provides functionalities such as controlling unnecessary complexity, and uncertainty estimation. 

The demonstration with a HEP use case clearly showed the capability of BNN as an unbinned regression tool, especially if several input variables with correlation are involved. This technique has already been used to analyze the data collected by LHC in 2010~\cite{SSpaper}. It has a very promising future for further applications, particularly for higher dimensional problems.

\section*{Acknowledgements}

We thank Andreas Hoecker, Joerg Stelzer, Peter Speckmayer, Jan Therhaag, Eckhard von Toerne and Helge Voss for their help in implementing this program into TMVA. We thank Song-Ming Wang and Zhili Weng for helpful discussions.  

Jiahang Zhong and Shih-Chang Lee are partially supported by the National Science Council, Taiwan under the contract number NSC99-2119-M-001-015.
\bibliographystyle{elsarticle-num}
\bibliography{nnfitting}

\end{document}